\documentclass[12pt]{article}
\usepackage{epsfig}
\usepackage{amsfonts}
\begin{document}
\begin{titlepage}
\begin{center}

{\Large Generalized statistical mechanics and fully developed
turbulence}

\vspace{2.cm} {\bf Christian Beck}

\vspace{1cm}

School of Mathematical Sciences, Queen Mary,
University of London, Mile End Road, London E1 4NS.

\vspace{5cm}

\end{center}

\abstract{The
statistical properties of fully developed hydrodynamic turbulence
can be successfully described
using methods from nonextensive statistical mechanics.
The predicted probability densities and scaling exponents precisely
coincide with what is
measured in various turbulence experiments.
As
a dynamical basis
for nonextensive behaviour we consider nonlinear Langevin
equations with fluctuating friction forces,
where Tsallis statistics can be proved rigorously.
}

\vspace{1.3cm}

\end{titlepage}

\small

\section{Introduction}

The formalism of
nonextensive statistical mechanics was
introduced by C. Tsallis in 1988 \cite{tsa1} and has been further
developed by many others \cite{abe}--\cite{tsa2}. Recently, it has been pointed out
in a number of papers that
the nonextensive approach is in particular
useful for the description of stochastic properties of
fully developed turbulent flows \cite{hydro}--\cite{12}. In this
paper we outline the main idea underlying this
new statistical mechanics approach to turbulence
und compare with various experimental measurements.

First we will summarize the basic idea of the nonextensive
formalism and point out some interesting connection to systems
with fluctuating temperature or fluctuating energy dissipation (i.e.\
nonequilibrium systems with a stationary state, see e.g.\
 \cite{eddie}). As an example
where everything can be calculated analytically, we will consider
suitable generalizations of Langevin equations with fluctuating
friction forces in section 3 \cite{benew1}. These underly the generalized
statistical mechanics in an analogous way as ordinary Langevin
equations underly ordinary statistical mechanics. The concrete
application to turbulent flows will be described in section 4.
Finally, in section 5 we compare with various experimental data in
turbulent flows. The agreement turns out to be very good, thus
indicating that methods from generalized statistical mechanics are
a useful tool for modeling turbulence. Fully developed turbulent
flows appear to effectively extremize the Tsallis entropies.

\section{Tsallis statistics and fluctuations}

In the nonextensive statistical mechanics approach, the usual
Shannon entropy $S_1= -\sum_i p_i \ln p_i$ is replaced by the more
general Tsallis entropies
\begin{equation}
S_q= \frac{1}{q-1} \left( 1- \sum_i p_i^q \right).
\end{equation}
The $S_q$ are related but different from the
Renyi entropies (see, e.g., \cite{BS}).
The $p_i$ are probabilities associated with the microstates $i$ of
the physical system, and $q\not= 1$ is a parameter, the `entropic
index'. The ordinary Shannon entropy $S_1$ is recovered
for $q \to 1$.

Extremizing $S_q$ under suitable constraints (see e.g. \cite{3} for
a review), just in an analogous way as it is done in ordinary
statistical mechanics, one formally obtains a $q$-generalized
version of the canonical distributions. The probabilities $p_i$
come out of the extremization procedure as
\begin{equation}
p_i= \frac{1}{Z_q} (1-(1-q)\beta E_i )^{\frac{1}{1-q}}, \label{tsa}
\end{equation}
where
\begin{equation}
Z_q= \sum_i (1-(1-q) \beta E_i)^{\frac{1}{1-q}}
\end{equation}
is the partition function, $\beta=1/(kT)$ is a suitable inverse
temperature variable, and the $E_i$ denote effective energy levels
of the microstates $i$ . In the limit $q \to 1$, ordinary
statistical mechanics is recovered, and eq.~(\ref{tsa}) reduces to
the ordinary Boltzmann factor $p_i \sim e^{-\beta E_i}$. Most
formulas of thermodynamics appear to have a simple generalization valid
for arbitrary $q$ \cite{3}.

Although this is certainly a beautiful mathematical framework, a
natural question is whether this type of
statistical mechanics is physically realized.
Tsallis' original suggestion was that it may have
physical applications for equilibrium systems with
long-range interactions (see \cite{latora} for a
recent update). Very recently, a slightly
different application was suggested.
It was pointed out that the
formalism is of particular physical relevance for nonequilibrium
systems with fluctuating temperature or fluctuating energy
dissipation \cite{beck01, benew1, wilk, benew2}. 
The basic idea how fluctuations can generate
nonextensive statistics is easily understood: Consider an
arbitrary Hamiltonian with energy levels $E_i$ of the
microstates. Using the integral representation of the gamma
function one can easily prove the following formula 
\begin{equation}
(1-(1-q)\beta_0 E_i)^{\frac{1}{1-q}}= \int_0^\infty
e^{-\beta E_i} f(\beta ) d\beta \label{marl},
\end{equation}
where
\begin{equation}
q-1=\frac{2}{n}
\end{equation}
and
\begin{equation}
f (\beta) = \frac{1}{\Gamma \left( \frac{n}{2} \right)} \left\{
\frac{n}{2\beta_0}\right\}^{\frac{n}{2}} \beta^{\frac{n}{2}-1}
\exp\left\{-\frac{n\beta}{2\beta_0} \right\} \label{fluc}
\end{equation}
is the probability density of the $\chi^2$ (or gamma) distribution
of order $n$. It arises in a natural way if $n$ independent
Gaussian random variables with average 0 are squared and summed
up.

Formula (\ref{marl}) is valid for arbitrary $E_i$ and
thus universal. The left-hand side of eq.~(\ref{marl}) is just the
generalized Boltzmann factor emerging out of nonextensive
statistical mechanics.
The right-hand side is a weighted average over
Boltzmann factors of ordinary statistical mechanics. If we
consider a nonequilibrim system with fluctuating temperature,
then nonextensive behaviour is a
consequence of integrating over all possible fluctuating inverse
temperatures $\beta$, provided $\beta$ is $\chi^2$-distributed.

The constant $\beta_0$ in eq.~(\ref{marl}) is the average of the
fluctuating $\beta$,
\begin{equation}
\langle \beta \rangle:=\int_0^\infty \beta f(\beta) d\beta =\beta_0
\end{equation}
and for the relative variance of the fluctuations one obtains
\begin{equation}
\frac{\langle \beta^2 \rangle -\langle \beta\rangle^2}{\langle
\beta\rangle^2}=\frac{2}{n}.\label{q-1}
\end{equation}



\section{Langevin equation with fluctuating friction forces}

We will now construct a concrete dynamics where Tsallis
statistics can be rigorously proved.

Consider a nonlinear Langevin equation
of the form
\begin{equation}
\dot{u}=-\gamma F(u)+\sigma L(t) \label{theree}
\end{equation}
where $F(u)=-\frac{\partial}{\partial u}V(u)$ is a nonlinear
forcing and $L(t)$ is Gaussian white noise. 
For the turbulence application, we are mainly interested
in power-law potentials of the form $V(u)=C|u|^{2\alpha}$. Given
some fixed ratio $\beta:=\gamma/\sigma^2$ of the parameters in the
stochastic differential equation
(\ref{theree}), this  
generates a process with the stationary probability density function
\begin{equation}
p(u|\beta)=\frac{\alpha}{\Gamma \left( \frac{1}{2\alpha}\right)}
\left( C\beta \right)^\frac{1}{2\alpha}\exp\left\{-
\beta C |u|^{2\alpha}\right\} . \label{there2}
\end{equation}
We now allow
the parameters $\gamma$ and $\sigma$ in
eq.~(\ref{theree}) to fluctuate as well.
Let us assume that either $\gamma$ or $\sigma$ or both
fluctuate in such a way that $\beta=\gamma/\sigma^2$ is $\chi^2$-distributed
with degree $n$. For example,
we can generate this $\chi^2$-distribution by yet another set
of Langevin equations, say
\begin{equation}
\dot X_j = -\Gamma X_j +L_j (t) \;\;\;\;\;\; (j=1, \ldots , n),
\end{equation}
and by defining
\begin{equation}
\beta =\frac{\gamma}{\sigma^2}:=\sum_{j=1}^{n} X_j^2 \label{Gauss}.
\end{equation}
On a long time scale this set of linear Langevin
equations clearly generates independent Gaussian random variables
$X_j$ with average 0 (provided $\Gamma =const$), hence $\beta$ is 
$\chi^2$-distributed
with density (\ref{fluc}).

Assume that the time scale on which $\beta$ fluctuates 
is very large, so that for a given $\beta$ local
equilibrium described by $p(u|\beta)$ is achieved in eq.~(\ref{theree}). In the
long-term run we have to take into account all fluctuating values
of $\beta$. This means we will ultimately observe the marginal
distribution $p(u):=\int p(u|\beta) f(\beta)d\beta$ being
generated by eq.~(\ref{theree}). This integral
can be performed exactly and one obtains
\begin{equation}
p(u)=\frac{1}{Z_q}\frac{1}{(1-(1-q)\tilde{\beta}C
|u|^{2\alpha})^{\frac{1}{q-1}}} \label{pu},
\end{equation}
where
\begin{equation}
Z_q^{-1}= \alpha \left( C(q-1)\tilde{\beta}\right)^{\frac{1}{2\alpha}}
\cdot \frac{\Gamma \left( \frac{1}{q-1}\right)}{ \Gamma
\left( \frac{1}{2\alpha} \right) \Gamma \left(
\frac{1}{q-1}-\frac{1}{2\alpha} \right)}
\end{equation}
and
\begin{eqnarray}
q&=&1+\frac{2\alpha}{\alpha n+1} \label{qn} \\
\tilde{\beta}&=&\frac{2\alpha}{1+2\alpha -q}\beta_0.
\end{eqnarray}

\section{Application to turbulent flows}

Let us now come to our main physical example, fully
developed turbulence. Let $u$ in eq.~(\ref{theree}) represent a
local velocity difference in a fully developed turbulent flow as
measured on a certain scale $r$. It is well known that the
dissipation fluctuates in turbulent flows, so our model makes
sense. The parameter $\beta =\gamma/\sigma^2$ of the stochastic
differential equation is an {\em a priori} unknown function of the
energy dissipation in the flow. Let us consider a model where
\begin{equation}
\beta = \frac{\gamma}{\sigma^2}=\epsilon_r\tau \Lambda .
\end{equation}
Here $\epsilon_r$ is the (fluctuating) energy dissipation rate
averaged over $r^3$ and $\tau$ is a typical time scale during
which energy is transferred. $\Lambda$ is a constant with
dimension $length^4/time^4$, its value is irrelevant for the
following. Both $\epsilon_r$ {\em and} $\tau$ can fluctuate, and
we assume that $\beta \sim \epsilon_r \tau $ is
$\chi^2$-distributed, thus ending up with the dynamical model of
the previous section.

We remind those readers not familiar with turbulence that the
averaged energy dissipation rate $\epsilon_r(\vec{x},t)$ in a
volume $V$ of size $r^3$ is defined as
\begin{equation}
\epsilon_r (\vec{x},t)= \frac{1}{r^3} \int_V \epsilon
(\vec{x}+\vec{r},t)d^3r,
\end{equation}
where
\begin{equation}
\epsilon (\vec{x},t)= \frac{1}{2}  \nu \sum_{i,j} \left(
\frac{\partial v_i}{\partial x_j}+\frac{\partial v_j}{\partial
x_i} \right)^2
\end{equation}
is the instantaneous dissipation, $\nu$ is the kinematic
viscosity, and $v_i$ is the velocity fluctuation (the deviation
from the mean velocity) in $i$ direction \cite{pope}. The physical
idea of considering the generalized Langevin equation
(\ref{theree}) with $\beta \sim
\epsilon_r \tau$ is that a test particle in the turbulent flow
moves for a while in a certain region with a given $\epsilon_r$,
then moves to another region with another $\epsilon_r$, and so on.
In this picture, we actually associate the fluctuations of $\beta$
with spatial fluctuations in a spatially extended system.


At the smallest scale of the turbulent flow, the Kolmogorov length scale
$\eta=(\nu^3/\epsilon)^{1/4}$, we can give a theoretical prediction of
the entropic index $q$. Here we can write
\begin{equation}
\beta = \epsilon \tau_\eta \Lambda = (\epsilon \nu)^{1/2}\Lambda
=(u_\eta)^2 \Lambda , \label{ghj}
\end{equation}
where $\tau_\eta:=(\nu /\epsilon)^{1/2}$ denotes the Kolmogorov
time and $u_\eta := \eta /\tau_\eta =(\nu \epsilon )^{1/4}$ is the
Kolmogorov velocity (see, e.g., \cite{pope}). If $\epsilon$
fluctuates then all these quantities fluctuate as well. Due to the
three spatial directions one can actually define 3 independently
fluctuating Kolmogorov velocities $u_\eta^j$, $j=1,2,3$ such that
$u_\eta=$ $|\vec{u}_\eta|=$ $\sqrt{\sum_{i=1}^3(u_\eta^j)^2}$. The
3 components $u_\eta^j$ describe the flow of energy into the three
different space directions. The simplest model assumption is that
these Kolmogorov velocities are Gaussian with average 0. This
means, we identify $X_j=\sqrt{\Lambda} u_\eta^j$ in
eq.~(\ref{Gauss}). Hence at the smallest scale the 3 space
dimensions lead to $n=3$ or, using eq.~(\ref{qn}), $q=
\frac{3}{2}$ if $\alpha = 1$.

Note that our approach allows us to view the
fluctuations of $\epsilon$ at the Kolmogorov scale in terms of a
(hypothetical) ordinary Brownian particle of mass $M$ that is
subjected to ordinary thermal noise of temperature $T$. The
fluctuating velocity $\vec{V}$ of this constructed Brownian
particle coincides with the fluctuating vector $\vec{u}_\eta$ of
Kolmogorov velocities. The Brownian particle just absorbs the
turbulent energy flow at the Kolmogorov scale. It bridges the gap
between thermal and macroscopic description.
Equipartition of energy yields
\begin{equation}
\frac{1}{2} M{\langle (u_\eta^1)^2 \rangle}=  \frac{1}{2} kT.
\end{equation}
Using eq.~(\ref{ghj}) we can estimate the mass of our
energy-absorbing Brownian particle as
\begin{equation}
M=\frac{3kT}{\nu^{1/2} \langle \epsilon^{1/2} \rangle }.
\end{equation}

\section{Comparison with experiments}

The quality of agreement of the generalized 
canonical distributions (\ref{pu})
with experimentally measured
densities is remarkable. To start with,
Fig.~1 shows probability densities of 
longitudinal velocity differences as measured by
Swinney et al. in a turbulent Taylor-Couette flow on
various scales $r$ \cite{BLS}. They are
very well fitted
by eq.~(\ref{pu}) choosing an $r$-dependent $q$
and
$\alpha=2-q$. 
In \cite{BLS},
in total 152 different densities were evaluated,
varying
the spatial scale $r$
and the Reynolds numbers $Re$. 
Previous fitting attempts using e.g.\ stretched
exponentials do not yield anything of comparable precision.

Next, Fig.~2 shows the results of measurements in another
turbulence experiment. Here we see a histogram of the acceleration $a$
of a test
particle advected by the turbulent flow (measurements of Lagrangian
turbulence by Bodenschatz et al. \cite{boden}). The acceleration has
been rescaled to variance 1. It can be
regarded as a velocity difference on the smallest time scale that
makes sense in turbulence, the Kolmogorov time scale. Hence our
consideration in the previous section predicts Tsallis statistics
with $q=3/2$ provided $\alpha=1$.
Formula (\ref{pu}) yields a very good fit of the measured
distributions (solid line in Fig.~2). 
Note that we do not use any fitting parameter in
this plot,
since $q=3/2$ is predicted, and variance 1 implies 
$\tilde{\beta}C = 4$. On the contrary, in their paper \cite{boden}
Bodenschatz et al. fit their data
with the density function 
\begin{equation}
p(x)=c_1\cdot exp \left\{ - \frac{x^2}{c_2+c_3 |x|^{c_4}}\right\}, \label{bod}
\end{equation}
i.e.\ a
somewhat arbitrary modification of a stretched exponential
with several parameters $c_i$. It
yields a fit of similar quality as our formula but requires much more
free parameters.
Moreover,
there is no theoretical
reason whatsoever for a density of the form (\ref{bod}), whereas
nonextensive statistical mechanics, leading
to eq.~(\ref{pu}), is a well-defined theoretical
concept distinguished by an extremization principle and
generalized Khinchin axioms \cite{abe}.

Also scaling exponents can be evaluated using
nonextensive methods. The scaling exponents $\zeta_m$ of moments
of radial velocity differences $u= v(x+r)-v(x)$ measured at
distance $r$ in fully developed turbulent flows have a long
history (see, e.g.,
\cite{bohr, frisch}). In the inertial range one observes
scaling behaviour of the form
\begin{equation}
\langle | u|^m \rangle \sim r^{\zeta_m}.
\end{equation}
Recently it has been suggested to
calculate the $\zeta_m$ using nonextensive methods
\cite{hydro, becks, ari1, ari3}. One approach is
based on an extension of the multifractal model of turbulence
\cite{ari1}--\cite{ari3},
another one on a combination of the approach described here 
with so-called
extended self-similarity (ESS) \cite{benzi}.
A typical result of the latter approach is shown in Fig.~3
(see \cite{becks} for more details). The
nonextensive model clearly reproduces the experimental data of
scaling exponents very well. This is, by the way, not even
surprising, because if the densities (\ref{pu}) perfectly coincide with
the experimental densities than the
moments must coincide as well. The moments of the generalized
canonical distribution (\ref{pu}) can be analytically evaluated,
they are given by
\begin{equation}
\langle |u|^m \rangle =\int_{-\infty}^{+\infty}|u|^m p(u)du
=\left\{ \frac{1}{(q-1)\tilde{\beta}C}\right\}^{\frac{m}{2\alpha}} \frac{
B\left( \frac{m+1}{2\alpha},
\frac{1}{q-1}-\frac{m+1}{2\alpha}\right)}{ B\left(
\frac{1}{2\alpha} , \frac{1}{q-1}-\frac{1}{2\alpha} \right) },
\label{there}
\end{equation}
where $B(x,y)$ denotes the beta function.

Finally, one may also try to understand the scale-dependence of
the entropic index $q(r)$,
which has been measured very precisely in \cite{BLS}. Nonextensive
statistical mechanics does not say anything on the
function $q(r)$. For the $r$-dependence 
one needs additional theoretical input. One
possibility was recently suggested in \cite{benew2}. The basic
idea is as follows. The
Tsallis entropies are non-extensive, i.e.\
for independent subsystems I and II composed to a single
system I+II one has \cite{tsa1, 3, abe3}
\begin{equation}
S_q^I +S_q^{II}+(1-q)S_q^IS_q^{II}=S_q^{I+II}.
\end{equation}
The observation, however, is
that it is possible to make the Tsallis entropies quasi-additive
by choosing different entropic indices at different scales. I.e.,
given a certain $q$ for two small independent subsystems I and II
we may choose another $q'$ for the larger, composed system I+II
such that
\begin{equation}
S_q^I +S_q^{II} =S_{q'}^{I+II}. \label{bf}
\end{equation}
This property is called {\em quasi-additivity}
\cite{benew2}. In turbulence, $q$
is close to 1, so that a perturbative expansion in $q-1$ makes
sense. The final result of a calculation in \cite{benew2} is that if
higher-order contributions in $q-1$ are neglected then
quasi-additivity implies a power law for $1/(q-1)$ as a function
of $r$, 
\begin{equation}
\frac{1}{q-1} \sim \left( \frac{r}{\eta}
\right)^\delta ,\label{scal}
\end{equation}
provided $r$ is large enough. The exponent $\delta$ is
non-universal, at least for finite Reynolds numbers.
Fig.~4 shows that the Swinney data indeed confirm a power
law.

To summarize, the nonextensive approach
yields a very useful new method to describe the most important
statistical properties of fully developed turbulence. It is
in very good agreement with experimental measurements.

\vspace{0.3cm}

{\bf Figure captions}

\vspace{0.2cm}

{\bf Fig.~1} Experimentally measured probability densities in a Taylor-Couette
flow as measured by Swinney et al. \cite{BLS} for $r= 23, 46, 93,
208, 399, 830 \eta$
and comparison with formula
(\ref{pu}) with $q= 1.150, 1.124, 1.106, 1.086$,
$1.066, 1.055$  ($\alpha=2-q$). The $j$-th density is shifted by
$-j$ units. All densities are rescaled to variance 1.

{\bf Fig.~2} Experimentally measured measured probability density
of the acceleration $x:=a/\langle a^2 \rangle^{1/2}$
in Lagrangian turbulence as measured by
Bodenschatz et al. \cite{boden} for 
Reynolds numbers $R_\lambda=200,690,970$
and comparison with formula $(\ref{pu})$
with $q=3/2$ and $\alpha=1$.

{\bf Fig.~3} Scaling exponents $\zeta_m$ as measured in five
different experiments \cite{benzi}--\cite{tabe} and comparison
with the nonextensive prediction of \cite{becks}.

{\bf Fig.~4} $(q-1)^{-1}$ as a function of $r$. The data points are the
$q$-values obtained from the  
the measured densities of Swinney et al. \cite{BLS}, the straight line
corresponds to a power law as predicted
by quasi-additivity. The observed exponent is $\delta=0.30$.

\vspace{0.5cm}

\epsfig{file=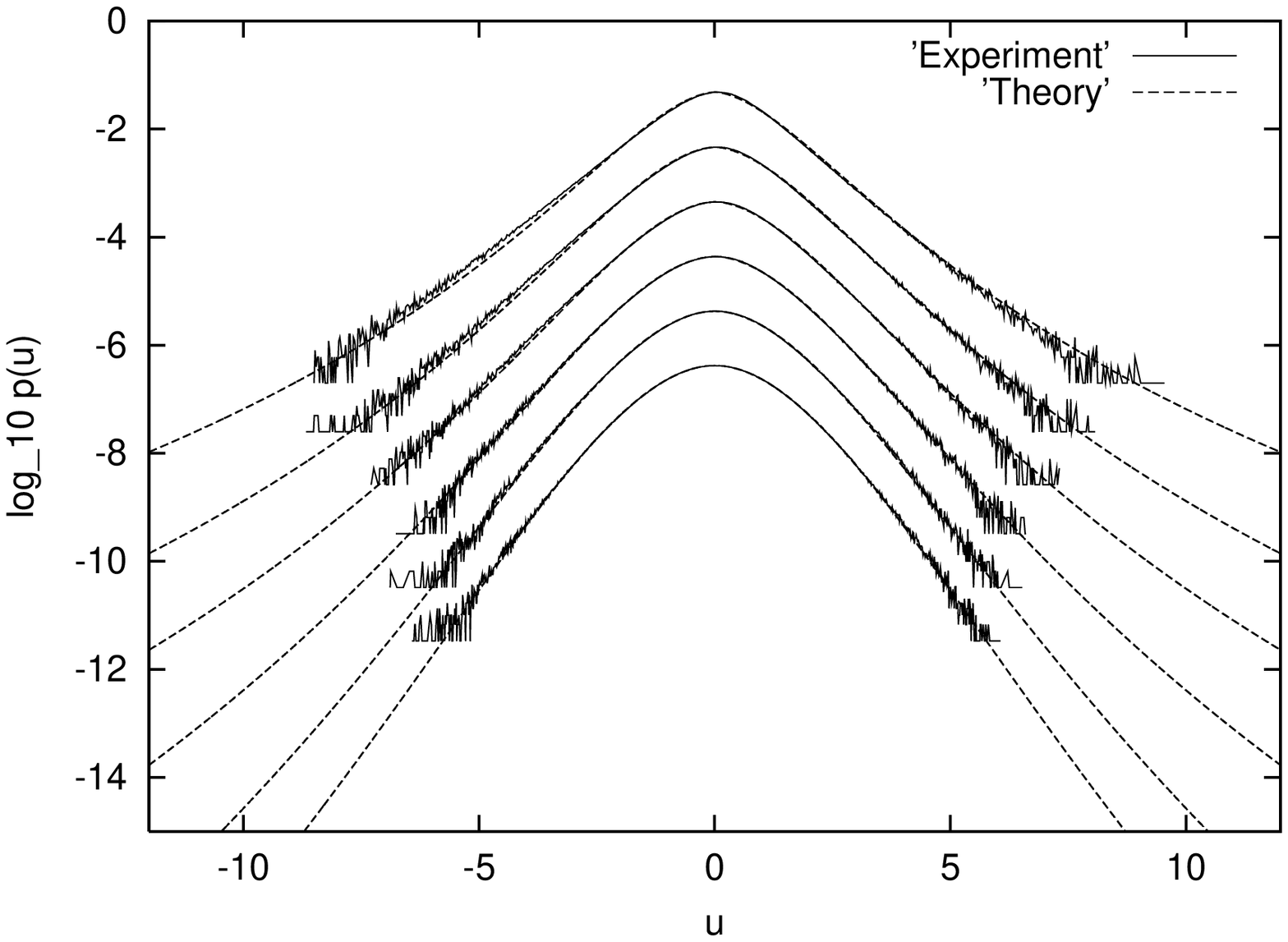, width=6cm, height=8cm, angle=-90.}

\epsfig{file=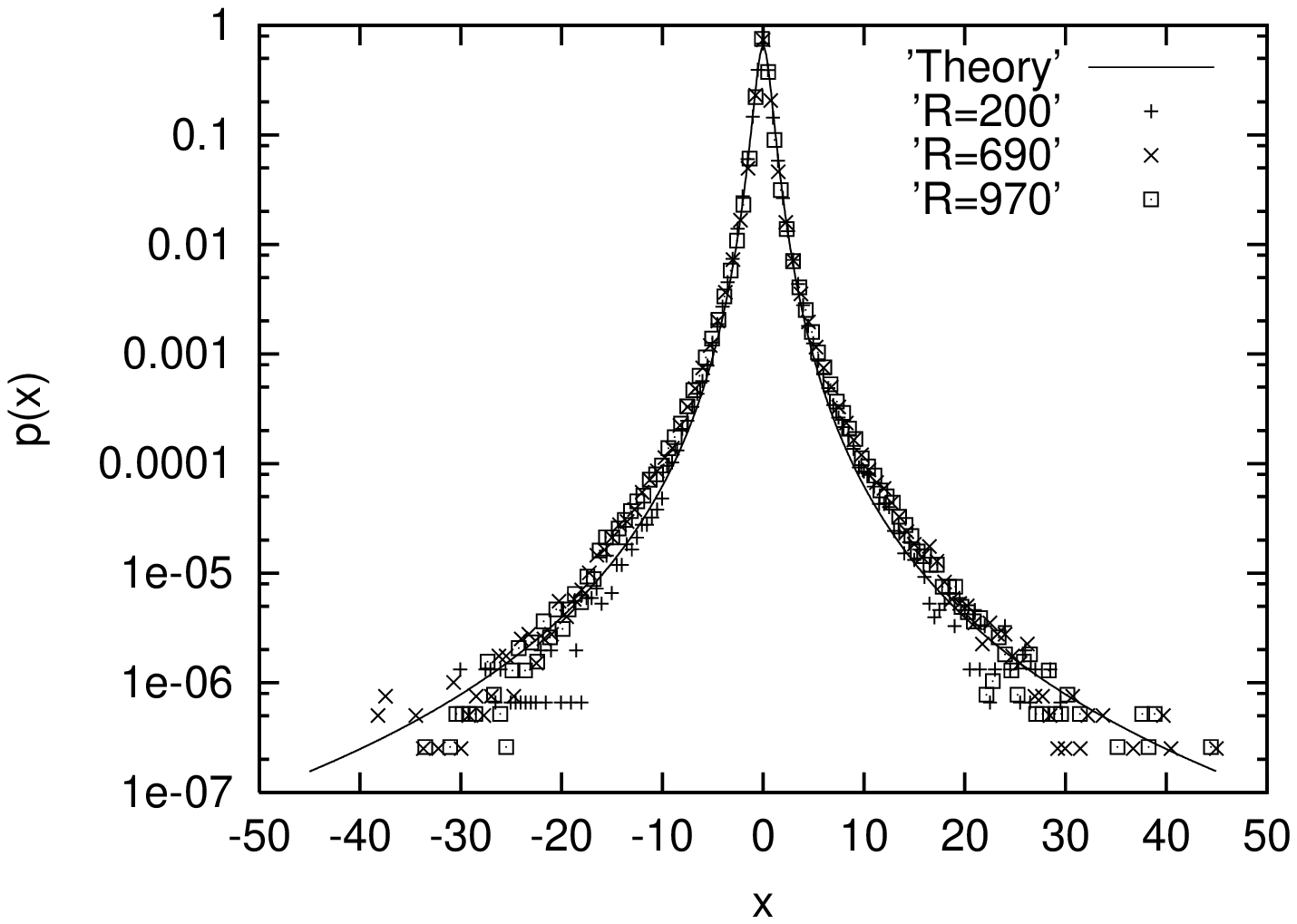, width=6cm, height=8cm, angle=-90.}

\epsfig{file=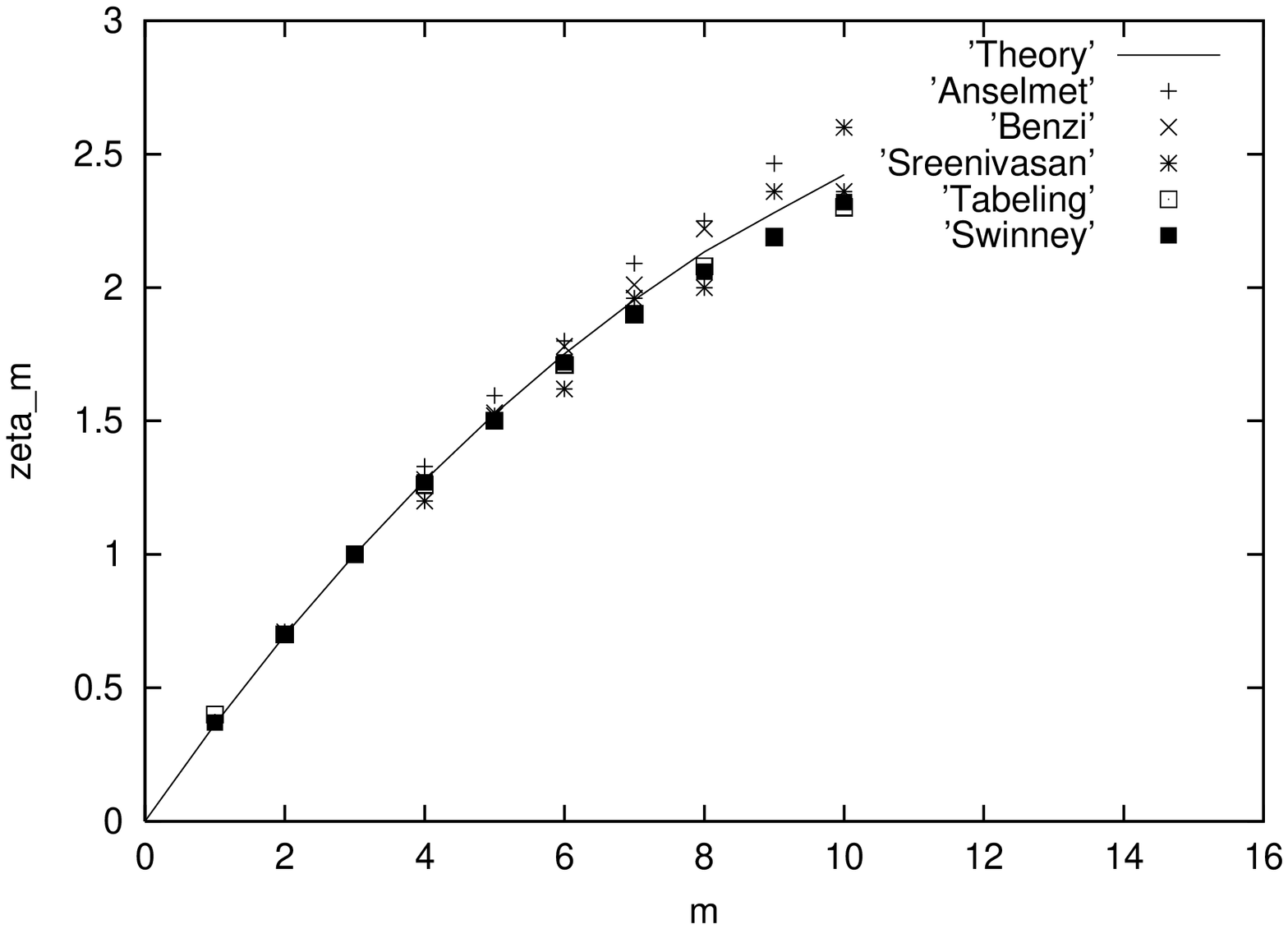, width=6cm, height=8cm, angle=-90.}

\epsfig{file=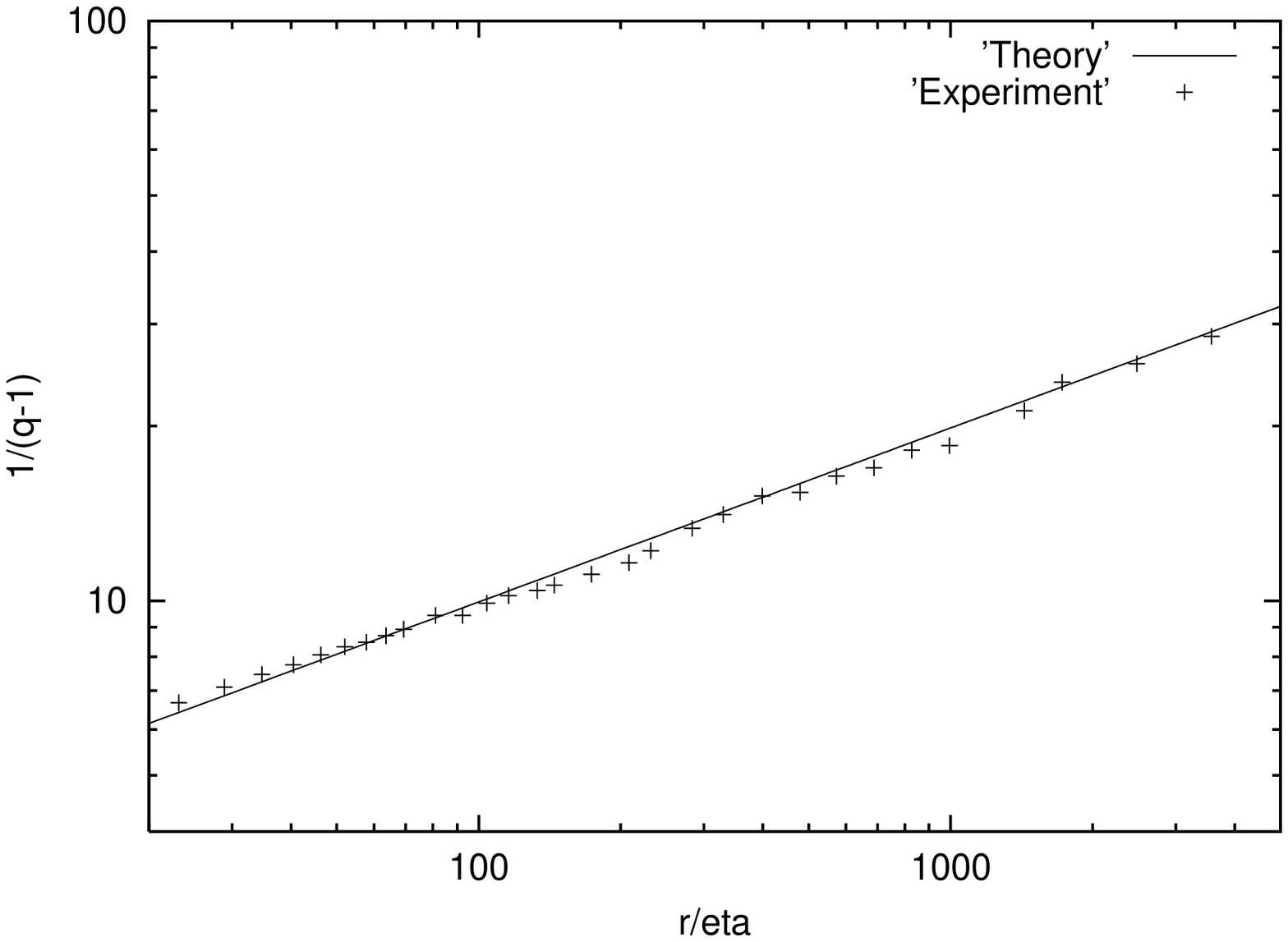, width=6cm, height=8cm, angle=-90.}

\end{document}